\documentclass{article}
\usepackage{spconf}
\usepackage{amsmath, amssymb, graphicx}
\usepackage{booktabs, array, multirow, makecell, adjustbox}
\usepackage{arydshln}          % for \hdashline / \cdashline
\usepackage[table]{xcolor}     % you use \textcolor{black!50}{...}
\usepackage{caption}
\usepackage[numbers]{natbib}
\usepackage{hyperref}
\title{Cross-Modal Bottleneck Fusion for Noise Robust Audio-Visual Speech Recognition}
\name{Seaone Ok$^{\star\dagger}$ \qquad Min Jun Choi$^{\star\ddagger}$ \qquad Eungbeom Kim$^\dagger$ \qquad Seungu Han$^\ddagger$ \qquad Kyogu Lee$^{\dagger \ddagger\S}$\thanks{$^{\star}$Equal contribution}}
\address{$^\dagger$IPAI,  $^\ddagger$Department of Intelligence and Information, $^\S$AIIS\\
Seoul National University, Republic of Korea}
\begin{document}
\ninept
\maketitle
\begin{abstract}
Audio-Visual Speech Recognition (AVSR) leverages both acoustic and visual cues to improve speech recognition under noisy conditions. A central question is how to design a fusion mechanism that allows the model to effectively exploit visual information when the audio signal is degraded, while maintaining strong performance on clean speech. We propose CoBRA (\textbf{C}r\textbf{o}ss-modal \textbf{B}ottleneck for \textbf{R}obust \textbf{A}VSR), a bottleneck-based fusion framework that introduces a compact set of learnable tokens to mediate cross-modal exchange. By regulating information flow through these tokens, the audio stream can reliably access essential visual cues even under adverse or out-of-domain noise. 
Despite limited training data, our model surpasses comparable baselines and remains competitive with large-scale systems through noise-adaptive fusion, demonstrating both efficiency and robustness. Ablation studies highlight that the depth of fusion is the most critical factor, underscoring its importance in designing robust AVSR systems.
\end{abstract}
%
% \vspace{2mm}
\section{Introduction}
\label{sec:intro}
% eb

\begin{figure}
    \centering
    \includegraphics[width=1.0\columnwidth]{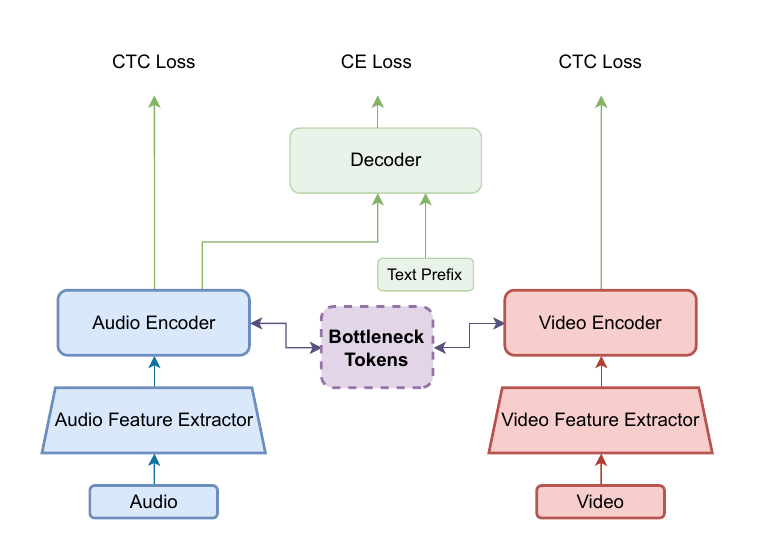}
    \caption{Overall architecture of the proposed model.}
    \vspace{-3mm}
    \label{fig:overall}
    \vspace{-1mm}
\end{figure}

% prev
Conventional automatic speech recognition (ASR) systems often suffer from severe performance degradation in noisy environments, where the audio signal alone is insufficient for reliable decoding. To overcome this limitation, audio-visual speech recognition (AVSR) integrates acoustic and visual modalities to improve robustness. Early studies showed that leveraging visual cues, such as lip movements, can significantly enhance recognition accuracy under challenging acoustic conditions~\cite{mroueh2015deepmultimodallearningaudiovisual, Afouras_2022, petridis2018endtoendaudiovisualspeechrecognition}. More recently, Transformer-based architectures~\cite{cm-seq2seq, cross-modal-AVSR, auto-avsr, whisper-flamingo} and large-scale pretraining~\cite{av-hubert, av-data2vec, haliassos2022jointly, haliassos2024braven} have substantially advanced AVSR. % , improving both accuracy and robustness. 
However, despite recent progress, two key limitations remain. First, cross-modal interactions are often suboptimal, depending on how and where modality fusion is applied. Second, previous methods based on cross-attention or naive concatenation introduce considerable computational overhead~\cite{zhao2024deep}. 
To address these limitations, we introduce CoBRA (\textbf{C}r\textbf{o}ss-modal \textbf{B}ottleneck for \textbf{R}obust \textbf{A}VSR), a bottleneck-based framework for AVSR. Our approach incorporates a small set of learnable bottleneck tokens on top of Conformer-based audio and visual encoders (see Figure.~\ref{fig:overall}) . Instead of direct attention between modalities, both streams interact exclusively through bottleneck tokens, which regulate multimodal information exchange. This mechanism, originally proposed for multimodal video classification~\cite{mbt}, encourages compact and efficient information sharing, while avoiding excessive computational cost.

Experiments on the LRS2 and LRS3 datasets show that bottleneck attention enables an effective AVSR framework that is both noise robust and data efficient, achieving strong performance with relatively less training data.
Under clean conditions, our model achieved WERs of 1.6\% and 2.8\% on LRS3 and LRS2, respectively. Under noisy conditions, CoBRA achieved a maximum of 7.42\% improvement on WER over the baseline. 
In addition, attention rollout~\cite{abnar2020quantifying} reveals its noise-adaptive behavior, as visual cues are increasingly leveraged under noisy conditions. Ablation studies illustrate that the position of the fusion layer is the most decisive factor for performance, with mid-level fusion providing the most reliable improvements. Overall, these findings establish CoBRA as an effective approach for noise robust AVSR.

% \vspace{-1mm}
\looseness=-1
\section{Related Works}
\label{sec:related work}
\subsection{End-to-End Conformer AVSR}
 A major milestone in AVSR was the introduction of end-to-end architectures built upon the Conformer, which combines convolutional modules with self-attention to effectively capture both local and global dependencies~\cite{conformer}. Ma et al.~\cite{cm-seq2seq} proposed an end-to-end audio-visual Conformer model that directly maps raw speech and visual input to transcriptions, trained with a hybrid CTC/attention objective~\cite{hybrid-ctc/attention}. Their framework uses ResNet-based front-ends instantiated for each modality, modality-specific Conformer encoders and a Transformer decoder, achieving strong performance on the LRS3 and LRS2 benchmarks.
\looseness=-1
\subsection{Fusion Methods in AVSR}
A range of fusion strategies have been explored in AVSR. The most straightforward approach is to concatenate features from both modalities at the input level, which is simple but often suffers from distribution mismatch. To address this, studies have proposed adversarial learning for modality-invariant representations~\cite{mir-gan} and attentive feature integration~\cite{attentive-fusion}. Another line of work combines decisions from separately trained unimodal systems at the prediction level, offering flexibility but limiting fine-grained cross-modal interaction~\cite{cm-seq2seq, auto-avsr, mlca-avsr}. More recently, attention-based methods~\cite{cross-modal-AVSR, mlca-avsr, cross-modal-AVSR2} explicitly model cross-modal alignment and improve robustness under noise, but often at the cost of high-dimensional representations and increased computation. Despite these advances, developing compact and noise-robust fusion mechanisms that fully exploit complementary audio-visual cues remains underexplored for robust AVSR.
\subsection{Attention Bottleneck Transformer}
Beyond conventional fusion strategies, the Bottleneck Transformer paradigm offers a systematic method to regulate how modalities exchange information. Nagrani et al.~\cite{mbt} proposed Multimodal Bottleneck Transformer (MBT), which inserts fusion bottlenecks at several layers in a Transformer backbone. MBT forces modality interactions to pass through a limited set of latent units, so that only essential information is shared while redundancy is suppressed. It achieves strong results on multimodal classification tasks, offering improvements in both accuracy and efficiency. While MBT establishes the effectiveness of fusion bottlenecks in video classification, its approach has yet to be adapted in depth for AVSR, where temporal resolution, sequence length, and decoding objectives introduce additional challenges. 

% \begin{figure}
%     \centering
%     \includegraphics[width=1.0\columnwidth]{figure/overall_architecture.drawio.pdf}
%     \caption{Overall architecture of the proposed model.}
%     \label{fig:overall}
% \end{figure}

\section{Method}
\label{sec:method}
\subsection{Overall architecture}
Our framework, as illustrated in Figure~\ref{fig:overall}, is based on a dual-stream design, where audio and visual inputs are first encoded separately and then integrated through bottleneck-based fusion. The overall pipeline consists of three main components: (i) modality-specific encoders, (ii) bottleneck tokens for cross-modal interaction, and (iii) a Transformer decoder for sequence prediction. \\
\noindent\textbf{Audio encoder} The audio stream follows the Conformer backbone~\cite{cm-seq2seq, conformer}, which processes log-Mel filter bank features through convolutional subsampling and stacked macaron-style Conformer blocks. \\
\noindent\textbf{Video encoder} The visual stream processes cropped mouth Region-Of-Interest (ROI) sequences using the 3D + 2D ResNet front-end followed by a Conformer stack with the same block topology as the audio stream. \\
\noindent\textbf{Decoder} The fused encoder outputs (Section~\ref{method: bottleneck}) are passed to a Transformer decoder, where the text sequence acts as the query, and the audio features serve as keys and values. During inference, the beam search integrates attention scores with CTC posteriors for prediction.
\begin{figure}
    \centering
    \includegraphics[width=1.0\columnwidth]{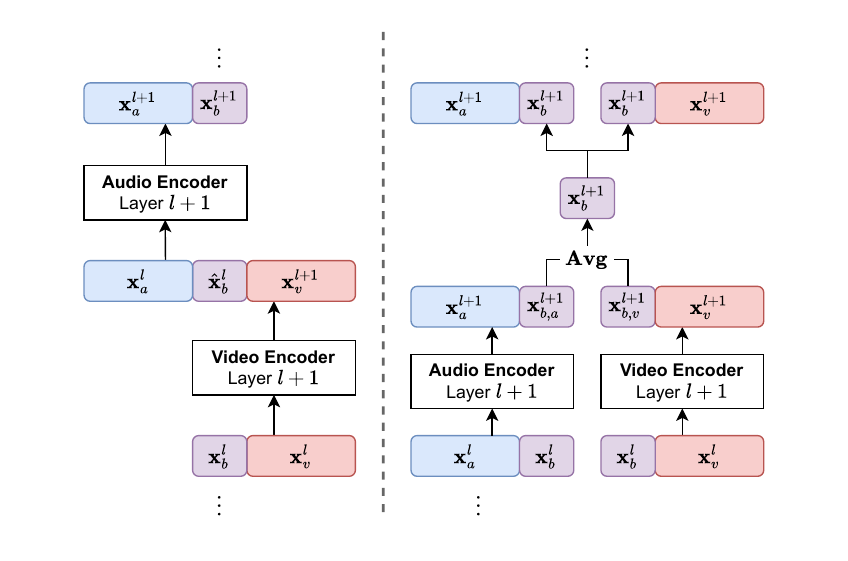}
    \captionsetup{skip=0pt}
    \caption{Illustration of bottleneck fusion strategies: \\ (left) sequential fusion and (right) mean fusion.}
    \label{fig:fusion}
    \vspace{-3mm}
\end{figure}

\subsection{Bottleneck-Based Modality Fusion}
\label{method: bottleneck}
Prior research~\cite{mbt} has shown that bottleneck embeddings can enhance modality fusion by constraining the flow of attention across modalities. Attention map visualizations in the earlier study suggest that bottleneck attention guides models toward task-relevant regions.
% helps models focus on regions most strongly associated with the output token. 
% This ability to filter out irrelevant signals is particularly desirable for AVSR
This property is especially desirable for AVSR, where noisy acoustic conditions often obscure modality-specific information. Building on these insights, we explore bottleneck attention as a strategy to improve noise robustness.
% Building on these insights, incorporating bottleneck attention into an AVSR framework may allow the model to adaptively emphasize informative cues under noisy conditions. To this end, we explore bottleneck attention as a potential strategy for improving noise robustness.
% Since Transformer complexity scales quadratically with sequence length, using a short bottleneck sequence enables more computationally efficient fusion compared to directly attending across all modality embeddings.

A schematic overview of the bottleneck update strategy is shown in Figure~\ref{fig:fusion}.
The bottleneck $\mathbf{x}_{b}\in \mathbb{R}^{F_b\times D}$ is parameterized as a set of learnable embeddings, where $F_b$ denotes the bottleneck sequence length.  For each modality $m$ at layer $l$, the embeddings are defined as $\mathbf{x}^{l}_{m} \in \mathbb{R}^{F_m\times D}$, with $F_m$ denoting the sequence length of modality $m$. The bottleneck is integrated into the encoder-only Transformer to facilitate cross-modal fusion, for which we investigate two update strategies: (a) sequential fusion and (b) mean fusion.

\noindent (a) sequential strategy:
\begin{subequations}
\begin{align}
\left[\hat{\mathbf{x}}^{l}_{b} || \mathbf{x}^{l+1}_v\right]
&=
\operatorname{Transformer}_v^l
\left(\left[\mathbf{x}^{l}_{b} || \mathbf{x}^{l}_v\right]\right)
\\
\left[\mathbf{x}^{l+1}_{b} || \mathbf{x}^{l+1}_a\right]
&=
\operatorname{Transformer}_a^l
\left(\left[\hat{\mathbf{x}}^{l}_{b} || \mathbf{x}^{l}_a\right]\right)
\end{align}
\end{subequations}

\noindent(b) mean strategy:
\begin{subequations}
\begin{align}
\left[\hat{\mathbf{x}}^{l+1}_{b,m} || \mathbf{x}^{l+1}_m\right]
&=
\operatorname{Transformer}_m^l
\left(\left[\mathbf{x}^{l}_{b} || \mathbf{x}^{l}_m\right]\right)
\\
\mathbf{x}^{l+1}_{b} &= \frac{1}{N(m)}\sum_{m}\left(\hat{\mathbf{x}}^{l+1}_{b,m}\right)
\end{align}
\end{subequations}
Here, $[~||~]$ denotes concatenation along the frame axis, $m$ indexes the modality, $N(m)$ is the total number of modalities and in this work $m \in \{{\text{audio}, \text{video}}\}$. Each modality is encoded independently up to layer $L_f$, where the bottleneck is then updated according to one of the strategies described above. The bottleneck embeddings are initialized from a Gaussian distribution and shared across the batch. By varying $L_f$, we systematically investigate how the depth of fusion affects the recognition performance, as discussed in Section~\ref{sec:ablation studies}.

\subsection{Training Objective}
\label{sec: training objective}
We adopt the hybrid CTC/Attention framework~\cite{hybrid-ctc/attention} and extend it by adding a video CTC loss alongside an audio CTC loss. Given audio/video input $\mathbf{x}$ and transcription $\mathbf{y}$, the training objective is
\begin{align}
\mathcal{L} = w \sum_{m}\log p_{\text{CTC},m}(\mathbf{y}|\mathbf{x}) + (1-w)\log p_{\text{CE}}(\mathbf{y}|\mathbf{x})
\end{align}
where $w$ denotes the relative weight between the CTC and attention-based components. This combined objective guides the encoder to learn temporally aligned representations while allowing the decoder to perform sequence-level modeling.

\section{Experimental Setup}
\label{sec:Experiments}
\subsection{Datasets}
We train and evaluate our framework on the LRS2~\cite{lrs2} and LRS3~\cite{lrs3} corpora, which are standard benchmarks for AVSR. The LRS2 dataset, extracted from BBC broadcasts, comprises approximately 224 hours of data, divided into a pre-train set (195 hours), a train-validation set (29 hours), and a test set (0.5 hours). The LRS3 dataset, built from TED and TEDx talks, contains roughly 438 hours of video and audio, partitioned into a pre-train set (407 hours), a train-validation set (30 hours), and a test set (1 hour). For both datasets, the official test sets are derived from videos that do not overlap with the training data.
\looseness=-1
\subsection{Implementation Details}
\noindent\textbf{Pre-processing} 
For visual stream, 96x96 mouth ROIs are cropped from each frame following a landmark-based pipeline~\cite{cm-seq2seq}, and normalized with the global mean-variance statistics of the training set. For audio stream, we perform the utterance level z-normalization. \\
\noindent\textbf{Data Augmentation}
We applied random cropping and time masking for video, and additive noise and time masking for audio inputs. During training, babble noise from NOISEX~\cite{noisex} is randomly mixed into the audio stream at varying SNR levels from -5~dB to ~20 dB. For evaluation, pink and white noise sampled from Speech Commands dataset~\cite{speech-commands} are added under controlled SNR conditions. \\
\noindent\textbf{Architecture} 
The audio and video stream uses a 1D and 3D + 2D ResNet front-end, respectively, followed by 12-layer Conformer encoders. Each encoder has an embedding dimension of 768, a feed-forward dimension of 3072, 12 attention heads, and a convolution kernel size of 31.
We introduce cross-modal fusion by concatenating 32 learnable bottleneck tokens with the embedding sequence. Fusion begins at the 4th Conformer layer of both encoders, and the fused representations are fed into a 6-layer Transformer decoder for text transcription. \\
\noindent\textbf{Training Setup}
All models are trained end-to-end using the loss function described in Section~\ref{sec: training objective}. We adopt a two-stage training strategy proposed by Ma et al.~\cite{auto-avsr}: (i) pretraining on short LRS3 utterances limited to 4 seconds in duration, with learning rate of 2.0e-4 for 50 epochs, and (ii) fine-tuning on the full LRS2 and LRS3 training sets with learning rate of 1.0e-3 for 75 epochs. We use AdamW optimizer ($\beta_1 = 0.9$ and $\beta_2 = 0.98$) with a cosine scheduler and 5-epoch warm-up. The global batch size is set to 57,600 frames. 

\begin{table}[]
\centering
\footnotesize                      
\setlength{\tabcolsep}{0.1pt}      
\caption{WER (\%) comparison on LRS2 and LRS3 evaluation sets. $\dagger$ fine-tuned with AV-HuBERT (video) and Whisper (audio).}
\begin{tabular}{
  >{\centering\arraybackslash}m{1.2cm} 
  >{\centering\arraybackslash}m{3.0cm}  
  >{\centering\arraybackslash}m{3.0cm}  
  >{\centering\arraybackslash}m{1.5cm} 
}
\toprule
\makecell{\bf Dataset} &
\makecell{\bf Method} &
\makecell{\bf Training Hours} &
\makecell{\bf WER (\%)} \\
\midrule
LRS2 
& Auto-AVSR~\cite{auto-avsr} & \makecell{3448} & 1.5 \\
& Whisper-Flamingo$^\dagger$~\cite{whisper-flamingo} & \makecell{1992} & 1.4 \\
& CM-seq2seq (baseline)~\cite{cm-seq2seq} & \makecell{381} & 3.7 \\
\cmidrule{2-4}
& CoBRA (Ours)       & \makecell{664} & 2.8 \\
\midrule
LRS3 
& AV-HuBERT~\cite{av-hubert} & \makecell{2192} & 1.4 \\
& Auto-AVSR~\cite{auto-avsr} & \makecell{3448} & 0.9 \\
& Whisper-Flamingo$^\dagger$~\cite{whisper-flamingo} & \makecell{3518} & 0.9 \\
& CM-seq2seq (baseline)~\cite{cm-seq2seq} & \makecell{596} & 2.3 \\
\cmidrule{2-4}
& CoBRA (Ours)       & \makecell{664} & 1.6 \\
\bottomrule
\end{tabular}

\label{tab:main results}
\vspace{-3mm}
\end{table}

\captionsetup[table]{skip=4pt}
% \sisetup{table-number-alignment = center, table-format=2.2}
\begin{table*}[t] 
\centering
\setlength{\tabcolsep}{5pt}
\footnotesize      
\caption{WER (\%) comparison on LRS3 between the baseline and our proposed model variants under three different noise conditions. \\ $F_b$ denotes the bottleneck length, $L_f$ indicates the position of the fusion layer, and S specifies the fusion strategy. \\ $\dagger$ We reproduced the audio-visual Conformer model of Ma et al.~\cite{cm-seq2seq} and report is as our baseline.}
\begin{adjustbox}{max width=\textwidth}
\begin{tabular}{cccccccccccccccccc} 
\toprule
\multirow{2}{*}{\bf Method} & \multirow{2}{*}{\bf Clean} &\multicolumn{5}{c}{\bf Babble} & \multicolumn{5}{c}{\bf Pink} & \multicolumn{5}{c}{\bf White}  \\ 
\cmidrule(l){3-7} \cmidrule(l){8-12}\cmidrule(l){13-17} &  & 12.5 & 7.5 & 2.5 & -2.5 & -7.5 & 12.5 & 7.5 & 2.5 & -2.5 & -7.5 & 12.5 & 7.5 & 2.5 & -2.5 & -7.5 \\
\midrule
{\bf Baseline$^{\dagger}$} & 2.30 & 2.41 & 2.83 & 4.17 & 8.22 & 18.58 & 2.97 & 3.58 & 5.79 & 12.25 & 27.51 & 4.12 & 6.91 & 12.97 & 26.53 & 41.63 \\
\midrule
{\bf CoBRA (Ours)} &  &  &  &  &  &  \\
% $\bf{F\!\!=\!\!4}$ $\bf{B\!\!=\!\!32}$ $\bf{S\!\!=\!\!seq}$ & 1.96 & 2.12 & \bf{2.22} & \bf{2.93} & 5.32 & 11.79 & \bf{2.25} & \bf{2.74} & \bf{4.23} & \bf{9.20} & 25.35 & \bf{3.37} & \bf{4.87} & \bf{9.21} & 22.69 & 40.66 \\
$\boldsymbol{L_f=4}$ $\boldsymbol{F_b=32}$ $\boldsymbol{S=\text{\textbf{seq}}}$ & 1.96 & 2.12 & \bf{2.22} & \bf{2.93} & 5.32 & 11.79 & \bf{2.25} & \bf{2.74} & \bf{4.23} & \bf{9.20} & 25.35 & \bf{3.37} & \bf{4.87} & \bf{9.21} & 22.69 & 40.66 \\
\midrule
% $\bf{F\!\!=\!\!0}$ \textcolor{black!50}{B=32} \textcolor{black!50}{S=seq} & 2.30	& 2.33 & 2.53 &	3.11 & \bf{4.67} & \bf{11.16} & 2.65 & 3.65 & 6.27 & 13.61 & 34.11 & 5.87 & 10.00 & 22.83 & 45.62 & 56.00 \\
$\boldsymbol{L_f=0}$ $F_b=32$ $S=\text{seq}$ & 2.30	& 2.33 & 2.53 &	3.11 & \bf{4.67} & \bf{11.16} & 2.65 & 3.65 & 6.27 & 13.61 & 34.11 & 5.87 & 10.00 & 22.83 & 45.62 & 56.00 \\
% $\bf{F\!\!=\!\!8}$ \textcolor{black!50}{B=32} \textcolor{black!50}{S=seq} & 2.28 & 2.50 & 3.01 & 3.89 & 6.67 & 15.21 & 2.76 & 3.32 & 5.20 & 11.32 & 30.41 & 4.14 & 6.37 & 11.98 & 29.66 & 46.42 \\
$\boldsymbol{L_f=8}$ $F_b=32$ $S=\text{seq}$ & 2.28 & 2.50 & 3.01 & 3.89 & 6.67 & 15.21 & 2.76 & 3.32 & 5.20 & 11.32 & 30.41 & 4.14 & 6.37 & 11.98 & 29.66 & 46.42 \\
\addlinespace[2pt]
\cdashline{1-17}[3pt/2pt]
\addlinespace[2pt]
% \textcolor{black!50}{F=4} $\bf{B\!\!=\!\!4}$  \textcolor{black!50}{S=seq} & 2.10 & 2.22 & 2.49 & 2.99 & 5.42 & 12.43 & 2.34 & 2.80 & 4.86 & 9.83 & 25.73 & 3.43 & 5.20 & 9.62 & 23.16 & 44.08 \\
$L_f=4$ ~$\boldsymbol{F_b=4}$~ $S=\text{seq}$ & 2.10 & 2.22 & 2.49 & 2.99 & 5.42 & 12.43 & 2.34 & 2.80 & 4.86 & 9.83 & 25.73 & 3.43 & 5.20 & 9.62 & 23.16 & 44.08 \\
% \textcolor{black!50}{F=4} $\bf{B\!\!=\!\!16}$ \textcolor{black!50}{S=seq} & \bf{1.95} & \bf{2.10} & 2.44 & 3.15 & 5.43 & 11.99 & 2.36 & 2.88 & 4.44 & 9.69 & 25.03 & 3.47 & 5.20 & 9.67 & 22.66 & \bf{39.61} \\
$L_f=4$ $\boldsymbol{F_b=16}$ $S=\text{seq}$ & \bf{1.95} & \bf{2.10} & 2.44 & 3.15 & 5.43 & 11.99 & 2.36 & 2.88 & 4.44 & 9.69 & 25.03 & 3.47 & 5.20 & 9.67 & 22.66 & \bf{39.61} \\
\addlinespace[2pt]
\cdashline{1-17}[3pt/2pt]
\addlinespace[2pt]
% \textcolor{black!50}{F=4} \textcolor{black!50}{B=32} $\bf{S\!\!=\!\!mean}$ & 2.01 & 2.15 & 2.40 & 3.24 & 5.66 &12.48 & 2.30 & 2.84 & 4.55 & 9.48 & \bf{24.19} & 3.46 & 5.15 & 9.78 & \bf{21.30} & 41.38 \\
$L_f=4$ $F_b=32$ $\boldsymbol{S=\text{\textbf{mean}}}$ & 2.01 & 2.15 & 2.40 & 3.24 & 5.66 &12.48 & 2.30 & 2.84 & 4.55 & 9.48 & \bf{24.19} & 3.46 & 5.15 & 9.78 & \bf{21.30} & 41.38 \\
\bottomrule
\end{tabular}
\end{adjustbox}

\label{tab:noise robustenss}
\end{table*}

\section{Results}
\label{sec:results}
\subsection{Main Results}
% Table~\ref{tab:main results} shows the WERs on LRS2 and LRS3 clean evaluation sets. Despite being trained on only 664 hours of data, our proposed model achieves competitive performance, indicating that bottleneck-based fusion enables efficient AVSR even under limited supervision.
For comparison, we report results against CM-seq2seq~\cite{cm-seq2seq} as a representative baseline trained on a comparable scale, while AV-HuBERT~\cite{av-hubert}, Auto-AVSR~\cite{auto-avsr} and Whisper-Flamingo~\cite{whisper-flamingo} serve as standard reference points among large-scale AVSR systems. No external language model was used in these comparisons to isolate the effect of the fusion architecture.
Table~\ref{tab:main results} summarizes the WER results on LRS2 and LRS3. On LRS3, CoBRA achieves a WER of 1.6\% using only 664 hours of training data. Despite substantially shorter training hours, our model performs on par with large-scale systems and yields consistent improvements over the baseline, underscoring its data efficiency. On LRS2, CoBRA achieves a WER of 2.8\%, which lags behind heavily trained systems, yet it consistently improves upon the baseline, confirming that our method remains competitive under limited-resource conditions.
%  WERs on the LRS2 and LRS3 clean evaluation sets. The main results are obtained from a model trained on the combined LRS2 and LRS3 datasets, with a total training duration of 664 hours. Despite using relatively less data compared to prior work trained with several times more supervision, our proposed model achieves competitive performance. These findings indicate that bottleneck-based fusion enables  data-efficient AVSR.
\subsection {Ablation studies}
\label{sec:ablation studies}
We conducted ablation experiments on the LRS3 dataset under clean and noisy conditions with babble, pink and white noise at different SNR levels ranging from +12.5dB to -7.5dB. As shown in Table~\ref{tab:noise robustenss}, the results confirm that bottleneck fusion improves robustness over the baseline. Notably, the performance gap between our model and the baseline widened as the SNR decreases, indicating stronger noise robustness. In particular, under the most challenging noisy setting at -7.5~dB SNR, our model showed a relative improvement of 40.0\% over the baseline on babble noise. Similar trends were observed for pink noise and white noise, demonstrating consistent robustness across noise types. Among different configurations, we found out that 32 tokens with fusion at the 4th layer and sequential strategy provide the most reliable performance. \\
\\
\noindent\textbf{Placement of Fusion Layer}
Mid-level fusion ($L_f=4$) consistently outperforms both the baseline and other fusion depths. It achieves the lowest WER on clean test sets while providing strong robustness under noisy conditions. For babble noise, which is included during training, mid-level fusion improves recognition accuracy across all SNR levels, showing a clear advantage under severe degradation. This robustness also generalizes to out-of-domain noise types such as pink and white, where other fusion depths suffer from significant performance drops. Overall, the results demonstrate that mid-level bottleneck fusion offers a reliable and noise-robust configuration for AVSR across both in- and out-of-domain conditions. \\
\noindent\textbf{\# of Bottleneck Tokens}
We compared different bottleneck sizes with the fusion layer fixed at $L_f=4$. Using only a small number of bottleneck tokens ($F_b=4$) consistently underperforms larger configurations, particularly under noisy conditions. Increasing the size to $F_b=16\text{ or }32$ yields substantial improvements across both clean and noisy test sets, with relative WER reductions exceeding 35\% on babble noise at ~–7.5 dB compared to the baseline. While $F_b=16$ and $F_b=32$ perform similarly in most cases, $F_b=32$ achieves offers the most stable configuration, showing consistently strong results across clean, in-domain, and out-of-domain conditions. \\
\noindent\textbf{Fusion Strategy}
Both sequential and mean update strategies outperform the baseline across clean and noisy conditions, but the difference between them is relatively small. In fact, while the sequential update shows slightly lower error rates on average, mean update performs better in certain adverse conditions. 
% Overall, the two strategies exhibit similar trends, which is consistent with previous research~\cite{mbt} that reported only marginal differences between alternative update mechanisms.
% Previous study~\cite{mbt} has reported only marginal performance differences across fusion update strategies. In AVSR as well, the difference between the two update strategies was modest; however, the sequential update consistently achieved slightly better overall performance.
% Previous studies have reported only marginal performance differences across fusion update strategies. In contrast, our experiments show that in the context of AVSR, the sequential update strategy provides more effective fusion.

\section{Analysis}
\subsection{Averaged Attention Rollout}
In this section, we analyze cross-modal influence using the attention rollout~\cite{abnar2020quantifying, dosovitskiy2020image} technique. The rollout value $\tilde{A}_l[i,j]$ represents the contribution of the $j$-th input embedding to the $i$-th output embedding. We extend this framework with bottleneck fusion to analyze how cross-modal influence between audio and video varies under different noise levels. 

\begin{figure}
    \vspace{+1mm}
    \centering    \includegraphics[width=1.0\columnwidth]{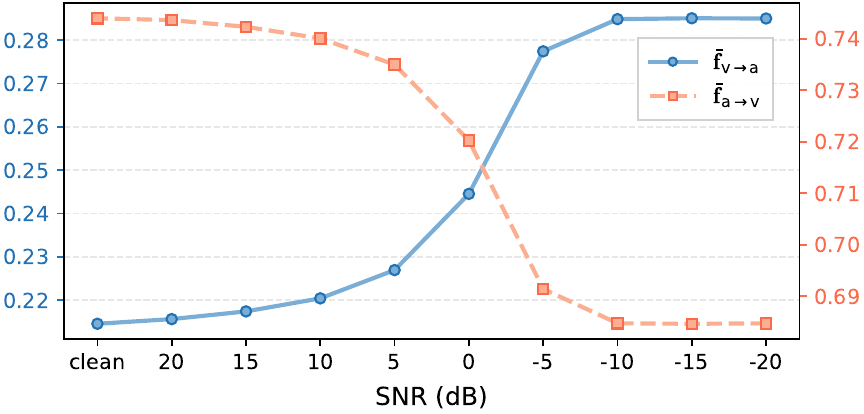}
    \captionsetup{skip=0pt}
    \caption{Variation of normalized cross-modal influence under different SNR conditions}
    \label{fig:SNR}
    \vspace{-1mm}
\end{figure}

From this formulation, the averaged influence of modality $m$ on modality $M$ is defined as
\begin{align}
f_{m \rightarrow M} &= \frac{1}{|\langle M \rangle| |\langle m \rangle|}\sum_{i \in \langle M \rangle}\sum_{j \in \langle m \rangle}\tilde{A}[i,j]
\end{align}
where $\langle m \rangle$ denotes the frame index set for modality $m$. To compare relative cross-modal contributions, we normalize by the total incoming mass, which includes both self-modal and cross-modal components. Specifically, for the audio and video streams, the normalized video-to-audio influence $\bar{f}_{v\to a}$ and audio-to-video influence $\bar{f}_{a\to v}$ are respectively defined as follows:
\begin{align}
\bar{f}_{v \to a} = \frac{f_{v \to a}}{f_{a \to a} + f_{v \to a}} \qquad \bar{f}_{a \to v} = \frac{f_{a \to v}}{f_{v \to v} + f_{a \to v}}
\end{align}
Values approaching 1 indicate strong cross-modal influence, whereas values near 0 suggest dominance of self-modal connections. Figure~\ref{fig:SNR} shows increasingly asymmetric cross-modal influence as noise intensifies, indicating a greater reliance on  visual information in forming audio representations. This behavior is consistent with known lip-reading ambiguities arising from the lack of one-to-one correspondence between phonemes and visemes, which can lead to omitted or confused sounds under severe noise.
% illustrates how video-to-audio normalized influence $\bar{f}_{v \to a}$ and audio-to-video normalized influence $\bar{f}_{a \to v}$ vary with the degree of noise injection. We hypothesize that under noisier conditions, video features become more informative, causing visual cues to play a more pronounced role in the final audio embedding. Variations in cross-modal interactions further demonstrate the model’s ability to exchange information in a noise-adaptive manner, as illustrated by representative failure cases under extreme noise. Representative failures such as “RACE IS A SOCIAL”→“RACIST AND SOCIAL” and “YOU’RE PLANNING”→“YOUR BRAIN IS” illustrate known ambiguities in lip reading that the lack of one-to-one correspondence between phonemes and visemes can lead to omitted plosives(/t/) or confusion between labial consonants(/b/,/p/). These cases contextualize the degradation observed in severe noise regi

\subsection{Bottleneck Fusion Configurations}
\textbf{Placement of Fusion Layer} Previous research~\cite{mbt} noted that early fusion can make attention overly permissive, leading to redundancy rather than effective information exchange. In our experiments, a similar trend is observed. Early and late fusion provide limited gains, whereas mid-level fusion ($L_f=4$) consistently achieves the best performance, offering a favorable balance between feature refinement and cross-modal interaction. % Early fusion (F=0) provides only limited benefits, which may be attributed to such redundancy effects. In contrast, with late fusion (F=8), the improvements are limited, suggesting that introducing cross-modal exchange too late leaves less room for effective interaction. Mid-level fusion (F=4) offers a favorable balance. This balance likely underlies the consistent performance gains observed with mid-level fusion in both clean and noisy conditions. % Each audio and visual features are sufficiently refined to avoid redundancy, yet still flexible enough to mediate adaptive information flow. 

\noindent\textbf{\# of Bottleneck Tokens}
Compared with the choice of fusion layer, the number of bottleneck tokens has a more limited impact on performance. However, we observe that overly short bottlenecks lead to reduced accuracy, whereas moderate to larger sizes result in more consistent improvements. Unlike video classification, AVSR requires modeling temporally varying speech under noise, making sufficient bottleneck capacity necessary for robust performance. 

% This observation contrasts with previous research~\cite{mbt}, which reported little effect of varying the number of tokens in video classification. A plausible explanation is that video classification operates on relatively uniform spatial patches, while AVSR must capture temporally variable speech signals that are also vulnerable to noise. Thus, a sufficiently large bottleneck capacity is required for the model to achieve robust performance in AVSR. 
% At the same time, we did not explore substantially larger capacities as the computational cost would become comparable to direct concatenation of video features, thereby undermining the efficiency advantages of the bottleneck design. 

\noindent\textbf{Computational Efficiency}
% Conventional fusion strategies concatenate audio and visual streams or apply cross-attention, resulting in quadratic complexity in doubled sequence length. In contrast, CoBRA performs fusion through B bottleneck tokens, leading to $O((F_m+F_b)^2)$. On LRS3 $(\text{L}\!\approx\!250, \text{B}\!=\!32)$, this corresponds to approximately $O(2\times1.27\text{L}^2)$, substantially lower than conventional designs, explaining the favorable efficiency-performance tradeoff.
Conventional fusion typically follows one of two designs: (i) concatenating the audio and visual frame sequences and applying full self-attention over the combined stream, or (ii) performing self-attention within each modality encoder and then exchanging information via cross-attention. Both approaches scale quadratically with sequence length, $O((2F_m)^2)$. In contrast, CoBRA performs fusion through $F_b$ bottleneck tokens, leading to $O\!\left(2(F_m+F_b)^2\right)$. Since $F_b \ll F_m$ in our setting, this substantially reduces attention computation compared to conventional strategies.

%\noindent\textbf{Fusion Strategy}
% In the ASR task, audio-only models substantially outperform video-only models, partly because visemes and phonemes do not correspond one-to-one. In many cases, several phonemes map to the same viseme, as they are visually indistinguishable when articulated~\cite{p2v_map}. This imbalance suggests that updating the bottleneck with equal weights across modalities can hinder effective information exchange. By contrast, the sequential update strategy, which allows the model to balance updates autonomously, yields more consistent performance across conditions. Notably, equal-weight updates provided minor gains in high-noise scenarios, where the audio signal carried limited information.
% We also observed differences across fusion update strategies, which appear to stem from an imbalance in information between modalities. In video classification, the performance gap between modalities is modest. In AVSR, however, audio-only and video-only models differ substantially. Notably, there is no one-to-one correspondence between visemes and phonemes: several phonemes map to the same viseme, as many phonemes are visually indistinguishable~\cite{p2v_map}.  Given this imbalance, we speculate that updating the bottleneck with equal weights across modalities may hinder effective fusion.

\section{Conclusion}
\label{sec:conclusion}
We proposed a bottleneck-based fusion framework for audio-visual speech recognition. Experimental results demonstrated consistent improvements over the baseline, validating its effectiveness of bottleneck tokens for cross-modal interaction. Ablation studies highlight that the choice of fusion position plays a decisive role in performance, underscoring its importance as a design factor. By enabling reliable integration of visual cues under noisy conditions, our approach enhances the robustness of AVSR systems, particularly when training data is limited. Future work could explore extending this approach to large-scale pretraining and modalities beyond audio and video, which may further enhance its generalization capability.

%\vfill
%\pagebreak

\section{Acknowledgements}
This work was partly supported by the National Research Foundation of Korea (NRF) grant funded by the Korea government (MSIT) [No. RS-2025-24683892, 50\%], Institute of Information \& communications Technology Planning \& Evaluation (IITP) grant funded by the Korea government(MSIT) [No. RS2022-II220641, 45\%], and [NO.RS-2021-II211343, 5\%] The GPUs were partly supported by the National IT Industry Promotion Agency (NIPA)'s high-performance computing support program in 2025.

\bibliographystyle{IEEEbib}
\bibliography{strings,refs}

\end{document}